\documentclass[11pt, a4paper]{article}

\usepackage{graphicx}
\usepackage{epstopdf}
\usepackage{amsmath}
\usepackage{amssymb}
\usepackage{amsfonts}
\usepackage{amsthm}
\usepackage[usenames]{color}
\usepackage{array}
\usepackage{verbatim}

\usepackage[
      colorlinks=true,
      linkcolor=blue,
      urlcolor=blue,
      filecolor=blue,
      citecolor=red,
      pdfstartview=FitV,
      pdftitle={},
      pdfauthor={},
      pdfsubject={},
      pdfkeywords={},
      pdfpagemode={},
      bookmarksopen=true
]{hyperref}

\usepackage{epsfig}

\usepackage{hyperref}

\def\beq{\begin{eqnarray}}
\def\eeq{\end{eqnarray}}

\def\be{\begin{equation}}
\def\ee{\end{equation}}
\def\bea{\begin{eqnarray}}
\def\eea{\end{eqnarray}}

\def\nn{\nonumber}

\numberwithin{equation}{section}

\begin{document}

\begin{centering}

  \vspace{0cm}

\textbf{\Large{
On Late Time Tails in an Extreme  Reissner-Nordstr\"om \\  \vspace{0.2cm} Black Hole:   Frequency Domain Analysis}}

 \vspace{0.8cm}

  {\large Srijit Bhattacharjee${}^1$, Bidisha Chakrabarty${}^{2}$, David D.~K.~Chow${}^{3}$\\ Partha Paul${}^{4,5}$, and Amitabh Virmani$^{4,5,6}$}

  \vspace{0.5cm}

\begin{minipage}{.9\textwidth}\small  \begin{center}
 ${}^{1}$Indian Institute of Information Technology, Allahabad \\
Devghat, Jhalwa, Uttar Pradesh 211015, India\\
  \vspace{0.5cm}
$^2$International Centre for Theoretical Sciences (ICTS), \\ Tata Institute of Fundamental Research, \\  Shivakote, 
Bengaluru 560 089, India \\
  \vspace{0.5cm}  
   ${}^{3}$Max Planck Institute for Gravitational Physics (Albert Einstein Institute),\\Am M\"{u}hlenberg 1, 14476 Potsdam-Golm, Germany\\
  \vspace{0.5cm}  
   ${}^{4}$Institute of Physics, Sachivalaya Marg, \\ Bhubaneswar, Odisha 751005, India  \\
  \vspace{0.5cm}
  ${}^{5}${Homi Bhabha National Institute, Training School Complex, \\ Anushakti Nagar, Mumbai 400085, India}\\
\vspace{4mm}
 ${}^{6}$Chennai Mathematical Institute, 
H1, SIPCOT IT Park, Siruseri  \\ Kelambakkam, Tamil Nadu 603103, India   \\

  \vspace{0.5cm}
{\tt srijitb@iiita.ac.in, bidisha.chakrabarty@icts.res.in, \\ david.chow@aei.mpg.de, pl.partha13@iopb.res.in, avirmani@cmi.ac.in}
\\ $ \, $ \\

\end{center}
\end{minipage}

\end{centering}

\begin{abstract}
In this brief note, we revisit the study of the leading order late time decay tails of massless scalar perturbations outside an extreme Reissner-Nordstr\"om  black hole. Previous authors have analysed this problem in the time domain; we analyse the problem in the frequency domain.  
We first consider initial perturbations with generic regular behaviour across the horizon on characteristic surfaces. For this set-up, we reproduce some of the previous results of Sela [arXiv:1510.06169] using Fourier methods. Next we consider related initial data on $t=\mbox{const}$ hypersurfaces, and present decay results at timelike infinity, near future null infinity, and near the future horizon. Along the way, using the $r_* \to -r_*$ inversion symmetry of the extreme Reissner-Nordstr\"om spacetime, we relate the higher multipole Aretakis and Newman-Penrose constants for a massless scalar in this background.  

\end{abstract}

\newpage
\tableofcontents

\setcounter{equation}{0}
\section{Introduction}

More than 45 years ago, Price \cite{Price:1971fb}, in his seminal analysis, showed that when a Schwarzschild black hole is perturbed by a massless scalar field, at late times the perturbation typically decays as an inverse power in the Schwarzschild coordinate $t$. Price's law has been rigorously proved in the mathematical general relativity literature by Dafermos and Rodnianski \cite{Dafermos, DafermosReview}.
This is a key result, as the problem of late time asymptotics for solutions to the wave equation
finds important applications in the study of black hole stability \cite{Dafermos:2008en, Dafermos:2010hd, Dafermos:2016uzj}  and the dynamics of black hole interiors \cite{Poisson:1990eh, Dafermos:2003wr, Marolf:2011dj}. The late time asymptotics to wave equations on extreme black holes have attracted exceptional interest in the last few years.  

The problem of late time decay of a scalar perturbation in 
 four-dimensional extreme Reissner-Nordstr\"om black hole was first analysed by Bi\v{c}\'{a}k \cite{Bicak}.  He observed that the effective potential for a massless scalar in an extreme Reissner-Nordstr\"om black hole has the same asymptotic form near the horizon as near infinity. Couch and Torrence \cite{CT} later showed that not only the effective potential has the same asymptotic form, it is in fact symmetric under $r_*$ going to $-r_*$, where $r_*$ is the tortoise coordinate for the extreme Reissner-Nordstr\"om  metric.  This surprising symmetry allows one to relate scattering dynamics near the horizon to the asymptotic region.  This symmetry adds several novel features to the late time dynamics of a massless scalar field in an extreme Reissner-Nordstr\"om black hole background compared to a Schwarzschild black hole. This richness is one of the reasons that several authors have studied this problem \cite{Blaksley:2007ak, Lucietti:2012xr, Ori, Sela}.
 
 Another reason the problem has attracted attention in the last few years is that Aretakis \cite{Aretakis:2011ha, Aretakis:2011hc, Aretakis:2012bm, Aretakis:2013dpa} has shown that a massless scalar has an instability at the future horizon of an extreme Reissner-Nordstr\"om black hole. More precisely, Aretakis showed that a massless scalar field decays at late time on and outside the future horizon, however, generically on the horizon its first radial derivative does not decay. This implies an instability. Since the first radial derivative of the scalar decays away from the horizon but not on the horizon, it follows that the second-derivative must blow up  at late times \emph{on the horizon.} The Aretakis instability was studied numerically in detail in \cite{Lucietti:2012xr}. They found excellent agreement with Aretakis' results. Using the Couch-Torrence symmetry, the Aretakis instability has been related to the similar growth in the behaviour of the derivatives of the massless scalar field at null infinity \cite{Bizon:2012we, Lucietti:2012xr}. Motivated by these developments,  more recently, Ori and Sela  \cite{Ori, Sela} have re-analysed analytically the problem of late time decay of scalar perturbations outside an extreme Reissner-Nordstr\"om black hole along the lines of Price's  analysis. These questions are currently  being explored in the mathematical general relativity literature  \cite{Angelopoulos:3,Angelopoulos:2,Angelopoulos:1} as well.
 
In this note we revisit this problem. While the previous authors \cite{Blaksley:2007ak, Lucietti:2012xr, Ori, Sela} have analysed the problem in the time domain, we analyse the problem in the frequency domain. Our analysis brings a different perspective. We reproduce and extend some of the previous results.  Using the Couch-Torrence symmetry, an initial data with regular behaviour across the horizon on the $v:=t+r_*=0$ surface can be mapped to an initial data on the $u:=t-r_* =0$ surface. Analysing this inverted initial data,
Sela \cite{Sela} has argued that there is a contribution to the late time tail in an extreme Reissner-Nordstr\"om background that is \emph{not} due to the curvature of the spacetime. This contribution  can be obtained from a flat space analysis of the inverted initial data. We first reproduce these results, including the exact coefficients, using rather simple Fourier methods.

Application of the  frequency domain Green's function technique requires knowing initial data on a $t=\mbox{const}$ surface. 
Obtaining a precise relationship between  characteristic initial data specified on $u=0$ and $v=0$ null surfaces and initial data specified on a $t=\mbox{const}$ Cauchy surface is a difficult problem. However, to the extent the above mentioned flat space analysis is valid, it can be easily done. We use solution of flat space wave equation to obtain the correct fall off on the $t=0$ surface near spatial infinity.  To this, we add a sub-leading term (slower fall-off)  proportional to the ``initial static moment'' and compute its contribution to the  late time tail. This contribution arises due to backscattering from the weakly curved asymptotic region of the spacetime. 

Along the way, using the Couch-Torrence symmetry \cite{CT}, we also relate higher multipole Aretakis and Newman-Penrose constants \cite{Newman:1968uj} for a massless scalar in an extreme Reissner-Nordstr\"om black hole background.  
 
 The rest of the paper is organised as follows. In section \ref{prelims} we review various interesting features that this problem has, namely, the Couch-Torrence symmetry and the construction of Aretakis and Newman-Penrose constants.
 In section \ref{late_time} we analyse the late time dynamics of a massless scalar in the frequency domain. We end with a brief summary and a discussion of open problems in section \ref{disc}.

\section{Massless scalar in 4d extreme Reissner-Nordstr\"om spacetime}
\label{prelims}
The massless scalar wave equation in a 4d extreme Reissner-Nordstr\"om spacetime has a number of rich features. In this section we review some of these features. Along the way, we relate higher multipole Aretakis and Newman-Penrose constants.

\subsection{Couch-Torrence discrete conformal isometry}
The extreme Reissner-Nordstr\"om solution has a discrete conformal isometry \cite{CT}. A similar discrete conformal isometry also exists for the extreme D1-D5 string and for the extreme D3 brane; see comments and references in \cite{Aharony:1999ti} and for recent discussions see \cite{Chow:2017hqe, Godazgar:2017igz}. We will use this symmetry in an important way in later sections, so we start with a brief review of this symmetry following \cite{Bizon:2012we, Lucietti:2012xr}. 
In static coordinates the extreme Reissner-Nordstr\"om metric takes the form,
\be
ds^2 = - \left( 1 - \frac{M}{r}\right)^2 dt^2 + \left( 1 - \frac{M}{r}\right)^{-2} dr^2 + r^2 d\Omega^2,
\ee
where $r$ is the area radial coordinate and $d\Omega^2$ is the line element of the unit 2-sphere. The Couch-Torrence symmetry is 
\be
\mathcal{T}: (t, r, \theta, \varphi) \to \left(t, M + \frac{M^2}{r-M}, \theta, \varphi \right). \label{CT}
\ee
It has number of interesting properties. 
It is an involution, i.e.,\
$
\mathcal{T}^2 = 1.
$
Its pull-back  on the Reissner-Nordstr\"om metric acts by a conformal transformation
\be
\mathcal{T}_*(g) =  \Omega^2 g, \qquad \mbox{where} \qquad \Omega = \frac{M}{r-M}.
\ee
On the tortoise coordinate $r_*$ defined by
\be
r_* = r - M + 2M \log \left( \frac{|r-M|}{M} \right) - \frac{M^2}{r-M},
\ee
so that $
\frac{dr_*}{dr} = \left( 1 - \frac{M}{r}\right)^{-2},
$
it acts as 
$
\mathcal{T}: r_* \to - r_*.
$
This last property implies that it interchanges the  ingoing and the outgoing Eddington-Finkelstein coordinates:
\begin{align}
\mbox{ingoing:}  \quad v&=t+r_*, & \mbox{outgoing:} \quad u&=t-r_*,&
\mathcal{T} :&  \quad u \leftrightarrow v. &
\end{align}

Since the Ricci scalar of the extreme Reissner-Nordstr\"om metric vanishes, the conformally covariant operator is simply the box operator:
\be
L_g = \Box_g - \frac{1}{6} R = \Box_g.
\ee
Recall that under a conformal transformation $\tilde g_{ab} = \omega^2 g_{ab}$, (see e.g., Wald's Appendix D, discussion around equation (D.13) \cite{Wald}),
\be
 L_{\omega^2 g} (\omega^{-1} \Phi) = \omega^{-3} L_g (\Phi) = \omega^{-3} \Box_g\Phi.
\ee
Moreover, from tensor transformation properties, it follows that
\be
L_{\mathcal{T}_*(g)} (\mathcal{T}_*(\Phi))= \mathcal{T}_*(L_g (\Phi)).
\ee
Combining the two in the following way,
it follows that if $\Box_g   \Phi = 0,$ then 
\be
0 =  \Box_g   \Phi = \mathcal{T}_*(L_g (\Phi))
= L_{\mathcal{T}_*(g)} (\mathcal{T}_*(\Phi)) = L_{\Omega^2 g} (\Omega^{-1} \Omega (\mathcal{T}_*(\Phi)) = \Omega^{-3} \Box_g (\Omega \mathcal{T}_* (\Phi)).
\ee
That is,  if $\Phi$ is a solution then,  
\be
\tilde \Phi = \Omega\mathcal{T}_*(\Phi) \label{scalar_map} 
\ee
is also a solution. We will use mapping \eqref{scalar_map} to map solutions near the horizon to solutions near future null infinity and vice versa. 

\subsection{Aretakis constants, Newman-Penrose constants, and initial static moments}

We now briefly review the construction of Aretakis and Newman-Penrose  constants in an extreme Reissner-Nordstr\"om background, and relate them via the Couch-Torrence symmetry \eqref{scalar_map}. 

Previous studies have related Aretakis and Newman-Penrose constants for $l=0$ modes \cite{Bizon:2012we, Lucietti:2012xr, Godazgar:2017igz}. To the best of our knowledge, details for the $l\neq0$ have not been written out. In this subsection we write out those details explicitly.

In ingoing Eddington-Finkelstein coordinates the extreme Reissner-Nordstr\"om metric is
\be
ds^2 = - \left( 1- \frac{M}{r} \right)^2 dv^2 + 2 dv dr + r^2 d\Omega^2.
\ee
Expanding the scalar in spherical harmonics in ingoing Eddington-Finkelstein coordinates as
\be
\Phi(v,r,\theta, \varphi) = \sum_{lm} \phi_{l}(v,r) Y_{lm}(\theta, \varphi),
\ee
we get equations for the mode functions $\phi_{l}(v,r)$,
\be
2 r \partial_v \partial_r (r \phi_{l}) + \partial_r ((r-M)^2 \partial_r \phi_{l}) - l(l+1) \phi_{l}= 0.
\ee
Applying $\partial_r^l$ on this equation, we see that 
\be
A_l [\phi_{l}] = \frac{M^l}{ (l+1)!} \partial_r^l [r \partial_r(r \phi_{l})] \bigg{|}_{r=M}
\ee
is conserved, i.e., independent of $v$ along the horizon. These constants are called Aretakis constants.
For a solution of the wave equation of the form near the horizon
\be
\phi_{l} (v, r) = \frac{1}{r}\sum_{k=0}^{\infty}c_k(v) \left(\frac{r}{M}-1\right)^k \label{near_horizon}
\ee
the Aretakis constants are \cite{Ori, Sela}
\be
A_l =  c_{l+1}  + \frac{l}{l+1}c_{l}. \label{Aretakis}
\ee
Note the factor of $\frac{1}{r}$ in equation \eqref{near_horizon}.

In outgoing Eddington-Finkelstein coordinates the extreme Reissner-Nordstr\"om metric is
\be
ds^2 = - \left( 1- \frac{M}{r} \right)^2 du^2 - 2 du dr + r^2 d\Omega^2.
\ee
Expanding the scalar in spherical harmonics in these coordinates as
\be
\Phi(u,r,\theta, \varphi) = \sum_{lm} \phi_{l}(u,r) Y_{lm}(\theta, \varphi),
\ee
we get equations for the mode functions
\be
-2 r \partial_u \partial_r (r \phi_{l}) + \partial_r ((r-M)^2 \partial_r \phi_{l}) - l(l+1) \phi_{l}= 0. \label{mode_eq}
\ee

We now construct the Newman-Penrose constants.  Consider the solution of the wave equation near infinity of the form
\be
\phi_l  (u,r)
= \frac{1}{r}\sum_{k=0}^{\infty} d_k(u) \bigg( \frac{M}{r} \bigg) ^k .
\label{inverse_power}
\ee 
Inserting this expansion into equation \eqref{mode_eq} and looking at successive inverse powers of $r$ gives equations that can be expressed concisely in terms of matrices, whose components are labelled by indices $i, j = 0, \ldots , l$.  We label the components of the vector $\mathbf{d}$ by $d_i$, $i = 0 , \ldots , l$, and the $(l + 1)$-dimensional vectors $\mathbf{d}_+$ and $\mathbf{c}_+$ have respective components $(d_+)_i = d_{i + 1}$ and $(c_+)_i = c_{i + 1}$.  We obtain
\be
M \mathsf{N}_l \dot{\mathbf{d}}_+ = [\tfrac{1}{2} l (l + 1) - \mathsf{P}_l] \mathbf{d} ,
\ee
where $\mathsf{N}_l$ is the diagonal matrix of natural numbers with components $(N_l)_{i j} = (i + 1) \delta_{i j}$ and $\mathsf{P}_l$ is a lower-triangular matrix with components $(P_l)_{i j} = \tfrac{1}{2} i (i + 1) \delta_{i j} - i^2 \delta_{i, j + 1} + \tfrac{1}{2} i (i - 1) \delta_{i, j + 2}$, and over-dots denote $u$-derivatives.  We can diagonalize $\mathsf{P}_l$ as
\be
\mathsf{P}_l = \mathsf{L}_l \mathsf{T}_l \mathsf{L}_l^{-1} ,
\label{diagonalt}
\ee
where $\mathsf{T}_l$ is the diagonal matrix of triangular numbers, $(T_l)_{i j} = \tfrac{1}{2} i (i + 1) \delta_{i j}$, and $\mathsf{L}_l$ is the lower-triangular Pascal matrix (see, e.g., \cite{edestr})
\be
(L_l)_{i j} = \binom{i}{j} ,
\ee
whose inverse is
\be
(L_l^{-1})_{i j} = (-1)^{i + j} \binom{i}{j} .
\ee
It follows that
\be
M \mathsf{L}_l^{-1} \mathsf{N}_l \dot{\mathbf{d}}_+ = [\tfrac{1}{2} l (l + 1) - \mathsf{T}_l] \mathsf{L}_l^{-1} \mathbf{d} ,
\label{dotd}
\ee
whose last component implies conservation of
\be
N_l :=  \frac{1}{l+1} \sum_{i = 1}^{l+1} (-1)^{l+i-1} i \binom{l}{i-1} d_i,
\ee
at null infinity, i.e., $\partial_u N_l =0$. These are examples of Newman-Penrose constants.

How are these constants related to Aretakis constants?  Recall that, applying the mapping \eqref{scalar_map}, we can construct a solution near null infinity from a given solution near the horizon. Let us apply this mapping on the solution of the form
 \eqref{near_horizon} to get
 \bea
 \phi_{l} &=& \frac{M}{r-M} \left(M + \frac{M^2}{r-M}\right)^{-1} \left( c_0(u) + c_1(u) \left(\frac{M}{r-M}\right) + c_2 (u) \left(\frac{M}{r-M} \right)^2 + \ldots \right) \\ 
 &=& \frac{1}{r} \left( c_0 + c_1 \frac{M}{r}  + (c_1 + c_2) \left(\frac{M}{r}\right)^2 + 
 (c_1+ 2 c_2  + c_3) \left(\frac{M}{r}\right)^3+
 \ldots \right).
 \eea
Expanding this solution in inverse powers of $r$, we find the coefficients in \eqref{inverse_power} to be
\be
\mathbf{d}_+ = \mathsf{L}_l \mathbf{c}_+ .
\ee
Then we have
\be
M \mathsf{Q}_l \dot{\mathbf{c}}_+ = [\tfrac{1}{2} l (l + 1) - \mathsf{T}_l] \mathsf{L}_l^{-1} \mathbf{d} ,
\label{dotc}
\ee
where
\be
\mathsf{Q}_l := \mathsf{L}_l^{-1} \mathsf{N}_l \mathsf{L}_l,
\label{diagonaln}
\ee
has components $(Q_l)_{i j} = (i + 1) \delta_{i j} + i \delta_{i - 1, j}$.  The Newman-Penrose constant $N_l$ arising from the last component of \eqref{dotd} and equivalently \eqref{dotc} is expressed in terms of $c_i$ as
\be
N_l = c_{l + 1} + \frac{l}{l + 1} c_l ,
\ee
which is nothing but  the Aretakis constant $A_l$, cf.~\eqref{Aretakis}.

The diagonalizations \eqref{diagonalt} and \eqref{diagonaln} are straightforwardly proved using induction on $l$, by checking that the columns of $\mathsf{L}_l$ and $\mathsf{L}_l^{-1}$ are eigenvectors of $\mathsf{P}_l$ and $\mathsf{Q}_l$ respectively.  As an explicit example, the Pascal matrices for $l = 4$ are

{\linespread{1}\selectfont \begin{align}
\mathsf{L}_4 & =
\begin{pmatrix}
1 & 0 & 0 & 0 & 0 \\
1 & 1 & 0 & 0 & 0 \\
1 & 2 & 1 & 0 & 0 \\
1 & 3 & 3 & 1 & 0 \\
1 & 4 & 6 & 4 & 1
\end{pmatrix} , &
\mathsf{L}_4^{-1} & =
\begin{pmatrix}
1 & 0 & 0 & 0 & 0 \\
-1 & 1 & 0 & 0 & 0 \\
1 & -2 & 1 & 0 & 0 \\
-1 & 3 & -3 & 1 & 0 \\
1 & -4 & 6 & -4 & 1
\end{pmatrix} ,
\end{align}
which diagonalize
\begin{align}
\mathsf{P}_4 & =
\begin{pmatrix}
0 & 0 & 0 & 0 & 0 \\
-1 & 1 & 0 & 0 & 0 \\
1 & -4 & 3 & 0 & 0 \\
0 & 3 & -9 & 6 & 0 \\
0 & 0 & 6 & -16 & 10
\end{pmatrix} , &
\mathsf{Q}_4 & = 
\begin{pmatrix}
1 & 0 & 0 & 0 & 0 \\
1 & 2 & 0 & 0 & 0 \\
0 & 2 & 3 & 0 & 0 \\
0 & 0 & 3 & 4 & 0 \\
0 & 0 & 0 & 4 & 5
\end{pmatrix} ,
\end{align}}

\noindent whose corresponding diagonal matrices $\mathsf{T}_4$ and $\mathsf{N}_4$ are simply their diagonal entries.  The Newman-Penrose constant is
\be
N_4 = \tfrac{1}{5} (\mathsf{L}_4^{-1} \mathsf{N}_4 \mathbf{d}_+)_4 = \tfrac{1}{5} (d_1 - 8 d_2 + 18 d_3 - 16 d_4 + 5 d_5) .
\ee

The term static moment is often used in the literature \cite{Price:1971fb} to discuss time independent solutions of the wave equations. In the static coordinates, the mode expansion,
\be
\Phi = \frac{1}{r} \sum_{lm} \psi_l(t,r) Y_{lm} (\theta, \varphi),
\ee
results in the equations
\be
[\partial_{r_*}^2 - \partial_t^2] \psi_{l} = V_l(r) \psi_l, \label{wave_eq}
\ee
with potential $V_l(r)$
\be
V_l(r) = \left( 1- \frac{M}{r}\right)^2 \left[ \frac{2M}{r^3}\left( 1- \frac{M}{r}\right) + \frac{l(l+1)}{r^2}\right].
\ee
This equation has two time independent solutions
\be
\psi_l = r (r-M)^l, \label{growing}
\ee
and
\be
\psi_l = \frac{r}{(r-M)^{l+1}}. \label{decaying}
\ee
Under the mapping \eqref{scalar_map} one static solution goes to the other (up to normalisation):
\be
\tilde \Phi = \Omega \,  \mathcal{T}_* \left[\Phi(t,r,\theta, \varphi)=(r-M)^l Y_{lm}\right] =\frac{M^{2l+1}}{(r-M)^{l+1}} Y_{lm}.
\ee

\section{Late time behavior of scalar perturbation}
\label{late_time}

We are now in position to analyse the problem of late time decay of scalar field outside the horizon in an extreme Reissner-Nordstr\"om background.

\subsection{Late time tails for non-compact initial data in flat space} 
We first reproduce some of the key results of Ori \cite{Ori} and Sela \cite{Sela} from a relatively simple Fourier analysis. In the next subsection we look at the contributions due to backscattering from the weakly curved asymptotic region.

\begin{figure}[t]
\centering
\begin{center}
\includegraphics[width=0.5\textwidth]{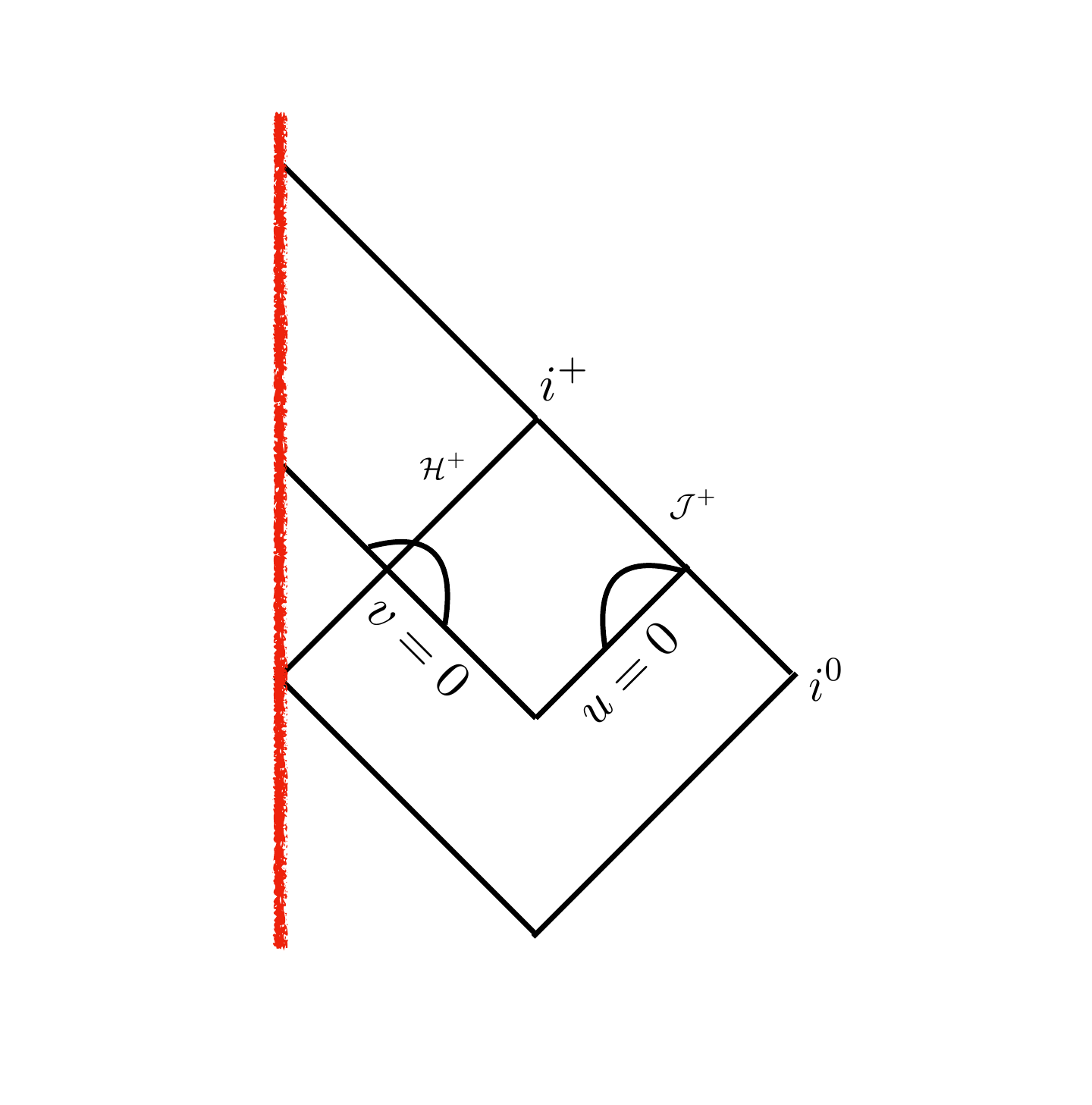}
\caption{Initial data for the characteristic initial value problem for a scalar field in an extreme Reissner-Nordstr\"om black hole background. The initial data is composed of two functions $\psi^v_l (v)= \psi_l(u=0, v)$ and $\psi^u_l (u) = \psi_l(u, v=0)$. }
\label{initial_data}
\end{center}
\end{figure}

To begin with, we  are interested in the characteristic initial value of the field $\psi_l$ specified at two intersecting null surfaces, $u=0$ and $v=0$, for equation \eqref{wave_eq}. The initial data is thus composed of two functions
\begin{align}
\psi^v_l (v) &= \psi_l(u=0, v), &
\psi^u_l (u) &= \psi_l(u, v=0). &
\end{align} 
See figure \ref{initial_data}. Due to the linearity of the problem, we can analyse the two functions $\psi^u_l (u)$ and $\psi^v_l (v)$ separately. More precisely, we can split the characteristic initial value problem into two parts: (i) non-vanishing data on the $u=0$ surface  $\psi^v_l (v) = \psi_l(u=0, v)$, along with vanishing data on the $v=0$ surface $\psi^u_l (u) = 0 $, (ii) non-vanishing data on the $v=0$ surface  $\psi^u_l (u) = \psi_l(u, v=0)$, along with vanishing data on the $u=0$ surface $\psi^v_l (v) = 0 $.
We can analyse the two parts separately and add the late time behaviour to obtain the final answer. In the following, this is how we will think of the evolution problem. For extreme Reissner-Nordstr\"om this logic has been employed by several authors in the past \cite{Blaksley:2007ak, Ori, Sela}.
 
Sela \cite{Sela} considered initial data of ``compact support'' --- an initial data for which the function $\psi^v_l (v)$ vanishes beyond certain value of $v$. In his analysis, the function $\psi_l^u(u)$ is taken to be supported near the event horizon $r=M$. Furthermore, this function is taken to admit a Taylor expansion near $r=M$ as
\be
\psi_l^u(u)=  c_0 + c_1  \left(\frac{r}{M}-1\right)+ c_2  \left(\frac{r}{M}-1\right)^2 + \ldots, \label{v_surface}
\ee
where $r$ is to be thought of as function of $u$ on the $v=0$ surface. 
We also take our initial data of this form for the function $ \psi^u(u)$. 
For the function $\psi^v_l (v)$ we consider a slightly more general behaviour than considered in \cite{Sela}. We allow for an initial static moment, i.e., as $r\to \infty$ the function $\psi^v_l (v)$ is taken to behave as 
\be
\psi^v_l (v) = \hat d_l \frac{R^l}{r^{l}} + \text{compactly supported data}, \label{u_surface}
\ee
where, now, $r$ is to be thought of as function of $v$ on the $u=0$ surface, and $R$ is an arbitrary scale we have introduced.  The coefficient $\hat d_{l}$ is called the static moment. 

For the initial data on the $u=0$ surface, it is believed that the late time tail arises due to backscattering from the weakly curved asymptotic region \cite{Price:1971fb, Klauder:1972je, Gundlach:1993tp}. The tail does not depend on the exact nature of the central object. For compactly supported initial data the solution at late times decays as $t^{-2l-3}$ and for initial data with initial static moment it decays as $\hat d_l t^{-2l-2}$ \cite{Price:1971fb}.

For the function $\psi^u(u)$, following \cite{Blaksley:2007ak, Ori, Sela}, we use the Couch-Torrence symmetry to map the problem from near the horizon to near infinity. The problem near infinity can be analysed again using the well developed techniques mentioned above. The map of the initial data is \eqref{CT}:
\be
\psi^v_l(v) = c_0 + c_1\left( \frac{M }{r -M} \right) + \ldots + c_l \left( \frac{M }{r -M} \right)^{l}   +
c_{l+1} \left( \frac{M }{r -M} \right)^{l+1} + \ldots
\ee
where now $r$ is to regarded as a function of $v$ along the $u=0$ surface. Expanding in powers of $r$ results in an expansion 
\be
\psi^v_l(v) = \hat c_0  + \hat c_1 \frac{R}{r} +\hat c_2 \frac{R^2}{r^2} + \ldots +  \hat c_l \frac{R^l}{r^l} + \hat c_{l+1}\frac{R^{l+1}}{r^{l+1}} + \ldots .
\ee
In this expansion there is a term that decays as the static moment.  There are terms that decay more slowly than the static moment and there are also terms that decay more quickly than the static moment. The coefficients  
$\hat c_{k}$ receive contributions from $c_{m}$, $m \le k$.

Again using linearity of the problem, the effective problem that we need to analyse is therefore, 
\be
\psi^v_l(v) = \hat c_0  + \hat c_1 \frac{R}{r} +\hat c_2 \frac{R^2}{r^2} + \ldots +  (\hat c_l + \hat d_l) \frac{R^l}{r^l} + \hat c_{l+1}\frac{R^{l+1}}{r^{l+1}} + \ldots + \text{compactly supported data}, \label{final_data_weak_field}
\ee
with $\psi^u_l(u)=0.$ Generically, if the Aretakis and the Newman-Penrose constants are non-zero, the coefficients of $\frac{1}{r^l}$ and $\frac{1}{r^{l+1}}$ would be non-zero in equation \eqref{final_data_weak_field}.

Sela \cite{Sela}, building upon the work of Barack \cite{Barack}, argued that for the initial data of the form \eqref{final_data_weak_field}, there is a contribution to the late time in an extreme Reissner-Nordstr\"om background that is not due to the curvature of the spacetime. The term
$
\hat c_{l+1}\frac{R^{l+1}}{r^{l+1}} 
$
in the expansion of the data \eqref{final_data_weak_field} results in a leading order tail as it disperses 
in flat space.\footnote{In some sense, this result is the ``Couch-Torrence dual'' of the results of section 4 of \cite{Lucietti:2012xr}, where they have obtained such tails from a purely $\text{AdS}_2 \times \text{S}^2$ analysis. $\text{AdS}_2 \times \text{S}^2$ is conformal to flat space, and the massless scalar wave equation is conformally invariant.} We reproduce Sela's results from a relatively simple Fourier analysis.

The Fourier transform of the field $\psi_l(t,r)$,
\be
\psi_l(\omega,r) = \int_{-\infty}^\infty e^{i\omega t} \psi_l(t,r) dt,
\ee  
satisfies the equation
\be
\label{radeq}
\left( -\omega^2 - \partial_r^2 + \frac{l(l+1)}{r^2} \right)\psi_l(\omega, r) = 0.
\ee
The general solution to this equation is 
\be
\psi_l(\omega,r) = A(\omega)\sqrt{r}J_{l+1/2}(\omega r) + B(\omega)\sqrt{r}Y_{l+1/2}(\omega r).
\ee
To obtain regular solutions at $ r=0 $ we must set $ B(\omega) = 0 $. Thus, we get 
\be
\label{soln2}
\psi_l(\omega,r) = A(\omega)\sqrt{r}J_{l+1/2}(\omega r) .
\ee
The solution in the time domain is simply the inverse Fourier transform,
\be
\label{soln3_1}
\psi_l(t,r) =  \frac{1}{2\pi}\sqrt{r} \int_{-\infty}^{\infty} A(\omega)J_{l+1/2}(\omega r) e^{-i\omega t} d\omega.
\ee
If we know the function $A(\omega)$, we can do this integral and would know the full solution for the field $\psi_l(t,r)$, in particular its late time behaviour. Due to linearity of the problem we can consider each term in the expansion \eqref{final_data_weak_field} separately.

To determine $A(\omega)$ corresponding to the $r^{-k}$ term, we use the fact that the initial data behaves as
$\hat c_{k}\frac{R^{k}}{r^{k}} $
on the $u=0$ surface. We make the ansatz $ A(\omega) = 2 \pi A_0 \ \omega^p $ to get from \eqref{soln3_1}
\begin{eqnarray}
\label{soln4}
 \psi_l(t,r) &=& A_0 \sqrt{r}\int_{-\infty}^{\infty} \omega^p J_{l+1/2}(\omega r) e^{-i\omega t} d\omega \\
 &=& 2 \ A_0 \ \sqrt{r} \ e^{i(p+l+1/2)\frac{\pi}{2}} \int_{0}^{\infty} \omega^p \ J_{l+1/2}( \omega r) \ \cos\left[ (p+l+1/2)\frac{\pi}{2} +\omega t \right] d\omega,
\end{eqnarray}
where we have used the appropriate symmetry property of $J_{l+1/2}( \omega r)$ under $\omega$ to $-\omega$ to convert the integral to one along the positive $\omega$ axis.
This last integral can be easily done using the identity (6.699-1) or (6.699-2)  of Gradshteyn and Ryzhik \cite{gradshteyn2007}.

Matching the resulting answer at $u=0$ with 
\be
\label{incond}
\psi_l(u=0, r) = \hat c_{k}\frac{R^{k}}{r^{k}},
\ee
gives 
\be 
p=k-1/2,
\ee 
and fixes the constant $A_0$. Substituting the constant $A_0$ in terms of $\hat c_{k}$ gives a final answer 
\bea
\label{soln9}
\psi_l(t,r) &=& -\frac{\hat{c}_k R^k 2^{k+1}\Gamma(k+1)}{\pi (2l+1)!!}  \sin(k\pi) \ \Gamma(l-k+1)\nn \\ & & \qquad \ r^{l+1} \ t^{-(k+l+1)}  \ F \left( \frac{l+k+2}{2}, \frac{l+k+1}{2}; l+\frac{3}{2}; \frac{r^2}{t^2} 
\right),
\eea
where $F(a,b;c;z)$ is the standard hypergeometric function. For $ k \leq l $ this expression vanishes due to the $\sin(k\pi)$ factor. However, for $ k \geq l+1 $, the $ \Gamma(l-k+1) $  factor develops a pole that exactly cancels with the zero of the $ \sin $ function and gives a finite result.  At timelike infinity, i.e., in the limit $ t \gg r$, \eqref{soln9} becomes
\be
\psi_l(t,r) \sim -\frac{\hat{c}_k R^k 2^{k+1}\Gamma(k+1)}{\pi (2l+1)!!} \  \sin(k\pi) \ \Gamma(l-k+1) \ 
r^{l+1} \ t^{-(k+l+1)}.
\ee
The leading contribution to the late time tail comes from $k=l+1$. We get
\be
\label{solution}
\psi(t,r | t \gg r) \sim 2\hat{c}_{l+1} R^{l+1}(-1)^{l+1} (4r)^{l+1} \frac{[(l+1)!]^2}{ (2l+2)!}  t^{-(2l+2)}.
\ee
This expression matches with Sela's equation (6.18), including the pre-factors. 

We can  use solution \eqref{solution_simple} to obtain the tail behaviour near future null infinity. In order to achieve the limit, we must take $r \to \infty$ together with $u := t-r$ finite, i.e., $u \ll r$. The leading contribution to the tail comes once again from $k=l+1$.
In this limit we find, using equation (9.131-2) of Gradshteyn and Ryzhik \cite{gradshteyn2007},
\be
\psi_l(t,r | u \ll r ) \sim 2^{l+2} \hat{c}_{l+1} R^{l+1}(-1)^{l+1} \frac{[(l+1)!]^2}{ (2l+2)!}  u^{-l-1}.
\ee
This expression matches with Sela's equation (6.11), including the pre-factors, provided we relate our $u$ to Sela's retarded time $u_s$: $u_s = u/2$. 

Setting $k=l+1$, equation \eqref{soln9} simplifies to 
\be
\label{solution_simple}
\psi_l(t,r) = 2\hat{c}_{l+1}R^{l+1} \frac{[(l+1)!]^2}{(2l+2)!} (-1)^{l+1} (4r)^{l+1} \left( 1- \frac{r^2}{t^2} \right)^{-l-1} t^{-2l-2}.
\ee
We can use the solution \eqref{solution_simple} to obtain the fall off  behaviour at spatial infinity,
\be
\psi_l(t=0,r) \sim 2^{2l+3} \hat{c}_{l+1} R^{l+1}  \frac{[(l+1)!]^2}{ (2l+2)!}  r^{-l-1}, \label{spatial0}
\ee
together with $\partial_t \psi_{l}(t=0,r) = 0.$

There are other contributions 
to the $t^{-(2l+2)} $ late time tail. They arise due to  backscattering from the curvature of spacetime.  
For initial data \eqref{final_data_weak_field}, these contributions come from  $r^{-k}$ terms for $k < l+1$. It is  expected that they should decay as 
\be
(\mbox{pre-factor}) \ M^{l+1-k} \hat c_k t^{-(2l+2)}.
\ee
We look at a related problem for $k=l$ in the next subsection. 

\subsection{Contributions due to asymptotic curvature of spacetime} 

\begin{figure}[t]
\centering
\begin{center}
\includegraphics[width=0.5\textwidth]{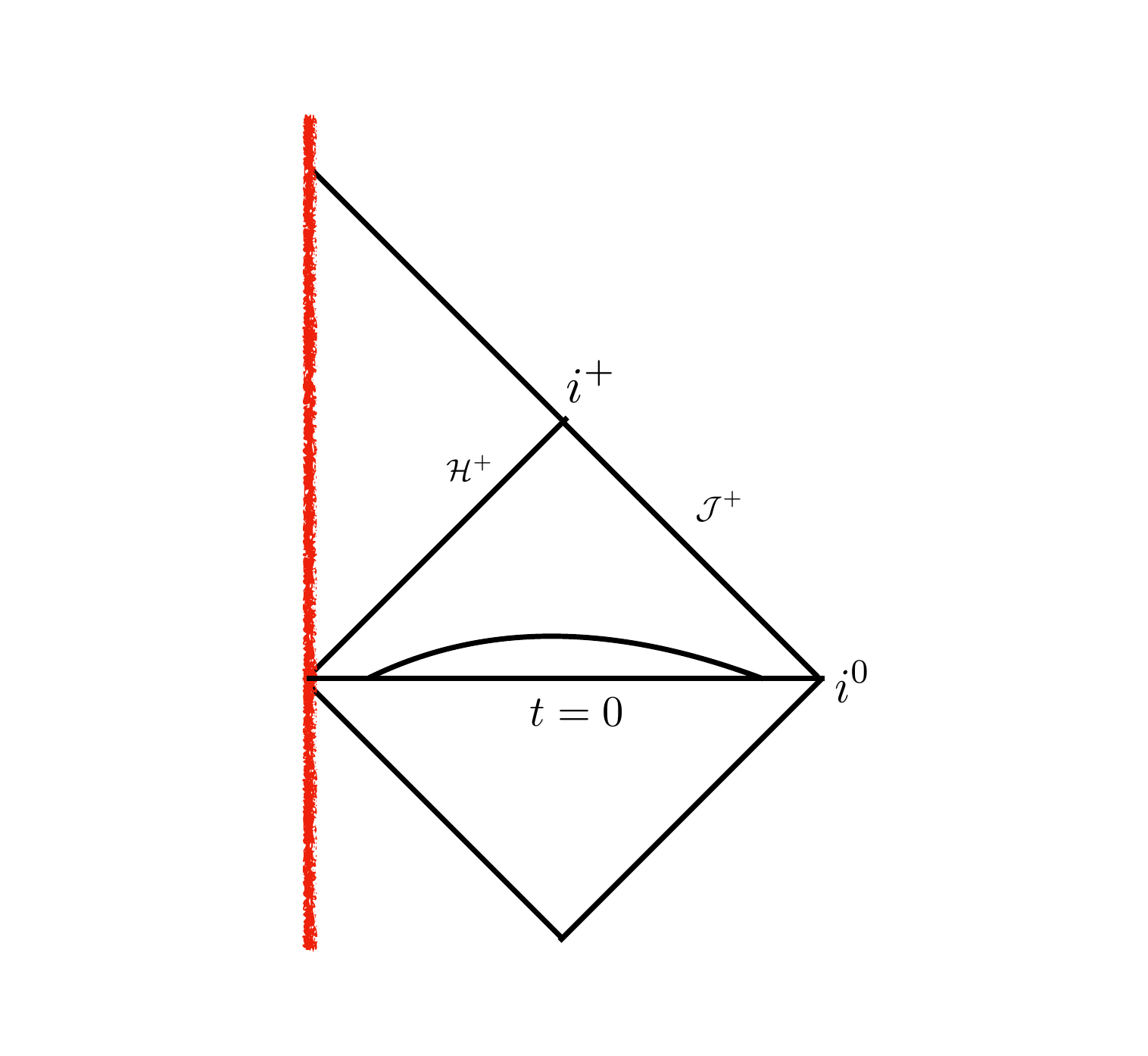}
\caption{Time symmetric initial data for scalar field in an extreme Reissner-Nordstr\"om black hole background. The initial data consists of a function specified on the $t=0$ surface.  $\partial_t\psi_l(t=0,r)$ is taken to be zero. }
\label{initial_data2}
\end{center}
\end{figure}

Equation \eqref{spatial0} can be interpreted as initial data on the $t=0$ surface in the extreme Reissner-Nordstr\"om background; see figure \ref{initial_data2}. We conclude that for an initial data on the $t=0$ surface with $r_*^{-l-1}$ decay near spatial infinity, there is a  contribution  to a $t^{-2l-2}$ tail at late times. This contribution is due to decay of a massless scalar field in flat space, not due to  backscattering from the curvature of the spacetime.   More precisely,  if 
\be
\psi_l(t=0,r) = \mu_{l+1}  R^{l+1} r_*^{-l-1}, \label{spatial}
\ee
then the late time tail is 
\be
\psi(t,r_* | t \gg r_* \gg M) \sim (-1)^{l+1}  2^{-l-1} \mu_{l+1}  R^{l+1}  (r_*)^{l+1} t^{-(2l+2)},
\ee
and
\be
\psi_l(t,r | u \ll r_* ) \sim (-1)^{l+1}  2^{-2 l-2}  \mu_{l+1}  R^{l+1}   u^{-l-1}.
\ee
From our discussion above, it is clear that such an initial data generically will have a non-zero Newman-Penrose constant, and its 
Couch-Torrence reflection will have a non-zero Aretakis constant. 

Now we address the question of if, in addition to \eqref{spatial}, there is an $r_*^{-l}$ term present at the $t=0$ surface near spatial infinity, how does it contribute to the late time tail? The contribution arises due to  backscattering from the curvature of spacetime. If the initial data has non-zero static moment, such a term would be present. 

 To compute this contribution, fortunately, we do not need to do much.  Since it is believed that the late time tail arises due to backscattering from the weakly curved asymptotic region \cite{Price:1971fb, Klauder:1972je, Gundlach:1993tp}, this computation is exactly the same as in the Schwarzschild background.

We very briefly review the Green's function approach to late time tails following \cite{Andersson:1996cm} and supplement it with a discussion for an extended source of the type:
\be
\psi_l(t=0,r) = \mu_{l}  R^{l}  r_*^{-l}. \label{spatial2}
\ee
 
The retarded Green's function for the wave operator appearing in \eqref{wave_eq} satisfies
\be
\left[\partial_t^2 - \partial_{r_*}^2  + V(r_*)\right] G(r_*, r_*'; t ) = \delta (t) \ \delta (r_*-r_*') 
\ee
with the boundary condition
\be
G(r_*, r_*'; t ) = 0, \qquad \mbox{for} \qquad t < 0. 
\ee
We are interested in analysing the Green's function in the frequency domain. Therefore,  we do a Fourier transform via
\be
\tilde G(r_*, r_*'; \omega )  = \int_0^\infty \, G(r_*, r_*'; t )  \, e^{i\omega t} \, dt .
\ee
The range of the $r_*$ coordinate for black hole spacetimes is $-\infty$ to $\infty$.  In the frequency domain the solutions to the wave equations we are interested in should satisfy outgoing boundary conditions at infinity,  and ingoing boundary conditions at the horizon. In terms of the $r_*$ coordinate, these become
\begin{align}
&\tilde \psi_l(r_*,\omega) \to e^{i\omega r_*} & & \mbox{as} & & r_* \to \infty,&\\
&\tilde \psi_l(r_*,\omega) \to e^{-i\omega r_*} & & \mbox{as} & &   r_* \to -\infty.&
\end{align}
The Fourier transform of the Green's function $\tilde G(r_*, r_*'; \omega )$ satisfies
\be
\left[ -\omega^2 - \partial_{r_*}^2 + V(r_*) \right] \tilde G(r_*, r_*'; \omega) = 0,
\ee
and is analytic in the upper half plane. Now recall that for a second order ODE with homogeneous boundary conditions, the Green's function can be uniquely constructed simply using two auxiliary functions
$f(r_*, \omega)$ and $g(r_*, \omega)$, where $f(r_*, \omega)$ satisfies the left boundary condition and $g(r_*, \omega)$ satisfies the right boundary condition.  We adopt normalisations such that
\begin{align}
&g(r_*, \omega)  \to  e^{i\omega r_*}& & \mbox{as} & & r_* \to \infty,& \\
&f(r_*, \omega)  \to e^{-i\omega r_*}& & \mbox{as} & & r_* \to -\infty.&
\end{align}
Then the Green's function is given by
\be
\tilde G(r_*,r_*'; \omega) = \begin{cases} 
\frac{f(r_*, \omega) g(r_*', \omega)}{W(\omega)}, & \quad \text{if } \quad r_* < r_*' \\
\frac{f(r_*', \omega) g(r_*, \omega)}{W(\omega)}, & \quad \text{if } \quad r_* > r_*'
  \end{cases}
\ee
where $W(\omega)$ is the Wronskian of the two solutions $f(r_*, \omega)$ and $g(r_*, \omega)$:
$
W(\omega) = g \partial_{r_*} f - f \partial_{r_*}g.
$
The Wronskian is independent of $r_*$. The late time tails come from the branch cut along the negative imaginary axis of the 
Green's function $\tilde G(r_*,r_*'; \omega) $ in the complex $\omega$ place \cite{Ching:1995tj}.

Andersson \cite{Andersson:1996cm} has presented a very clear computation of the branch cut of the Green's function in the low-frequency asymptotic expansion using some results from \cite{Futterman}. Instead of reviewing those details here, we simply write equation (40) of that reference (which has a typo of an overall minus sign)  \cite{Andersson:1996cm}:
\be
G^C(r_*,r_*',t) = - 2 \pi  i M \sqrt{r_* r_*'} \int_{0}^{-i \infty} \omega \ J_{l+1/2}(\omega r_*) \ J_{l+1/2}(\omega r_*') \ e^{-i \omega t}d\omega.
\label{cut}
\ee
The late time solution using this Green's function is simply (see e.g., equation (7.3.5) of \cite{MorseFeshbach} or \cite{Ching:1995tj})
\be
\label{soln1}
\psi^C_l(r_*,t) = \int_0^{\infty} G^C(r_*,r_*^{\prime},t)  \ \partial_t\psi_0(r_*^{\prime},0) \ dr_*^{\prime} - \int_0^{\infty}  \partial_t G^C(r_*,r_*^{\prime},t) \ \psi_0(r_*^{\prime})  \ dr_*^{\prime} , \, 
\ee
where we have implicitly used the fact that the leading contribution only comes from the asymptotic region, and our non-compact initial data has support only in the $r_* \gg M$ asymptotic region.

Inserting initial data  
\eqref{spatial2} together with
$ \partial_t\psi_0(r_*,0) = 0 $ in equation \eqref{soln1}, we get
\be\label{soln3}
\psi_l(r_*,t) =  2 \pi \mu_l R^l M \sqrt{r_*} \int_0^{ -i \infty}  d\omega \, \omega^2 \ e^{-i \omega t} \ J_{l+1/2}(\omega  r_*) \int_{0}^\infty dr^\prime_* \, {r^\prime_*}^{-l+1/2} \ J_{l+1/2}( \omega r^\prime_*).
\ee
We can evaluate the second integral in equation \eqref{soln3} using identity (6.561-14) of Gradshteyn and Ryzhik \cite{gradshteyn2007} to get
\bea
\psi_l(r_*,t) &=&   \frac{2 \sqrt{2 \pi}}{(2l-1)!!}  \mu_l R^l M \sqrt{r_*} \int_0^{ -i \infty}  \, \omega^{l+1/2} \ e^{-i \omega t} \ J_{l+1/2}(\omega  r_*) \, d\omega .
\eea
To compute the tail at timelike infinity, we approximate $ \omega r_* \ll 1$ to get
\bea
\psi_l (t,r_* | t \gg r_* \gg M)   &\sim&  \frac{4 \mu_l R^l M r_*^{l+1}}{(2l-1)!!(2l+1)!!}  \int_0^{- i \infty} \ \omega^{2l+1} \ e^{-i\omega t} \ d\omega\\
& = &  (-1)^{l+1} 4 \mu_l R^l M  \frac{ (2l)!!}{(2l-1)!!} r_*^{l+1} t^{-2l-2}. \label{main1}
\eea
This expression can be compared with equation  (69) of reference \cite{Leaver:1986gd} and equations IV-1 and IV-2 of reference \cite{Moncrief}. In those papers, computations are done differently, and in different contexts. 

To compute the tail near null infinity, we approximate $ \omega r_* \gg 1$. A similar calculation then  gives 
\be
 \psi_l (t,r_*) \sim (-1)^{l+1} 2 \mu_l R^l M \frac{ l!}{(2l-1)!!}  u^{-l-1}. \label{main2}
\ee
This expression can be compared with  equation  (68) of reference \cite{Leaver:1986gd}. 

Equations \eqref{main1} and \eqref{main2} are contributions  proportional to $\mu_l M$ to the late time tails in an extreme Reissner-Nordstr\"om black hole background.

\section{Discussion}
\label{disc}

In this note we have revisited the study of the leading order late time decay tails for massless scalar perturbations outside an extreme Reissner-Nordstr\"om black hole. While previous studies have analysed this problem in the time domain, we analysed the problem in the frequency domain.   A systematic time domain analysis was reported by Sela \cite{Sela}.\footnote{For electromagnetic and gravitational perturbations see \cite{Lucietti:2012xr, Sela:2016}.} Sela's analysis is quite involved. 
The merit of our work lies in its simplicity. 
We are able to obtain most of the key results of Sela's analysis, including all pre-factors, using rather straightforward Fourier methods.

We find that initial perturbations with generic regular behaviour across the horizon decays at late times as $t^{-2l-2}$ near timelike infinity $(t \gg r_*)$. It decays as $u^{-l-1}$ near future null infinity. The inversion map \eqref{scalar_map} maps the decay behaviour near future null infinity to the decay behaviour $v^{-l-1}$ near the horizon.

For initial data  of the form \eqref{final_data_weak_field} at the $u=0$ surface, there are other contributions 
to the $t^{-(2l+2)} $ late time tail. They arise due to  backscattering from the curvature of spacetime, from  terms $r^{-k}$ for $k < l+1$. These contributions should go as 
\be
\mbox{(pre-factor)} \ M^{l+1-k} \hat c_k t^{-(2l+2)}.
\ee
We have not addressed these contributions in this note. For $k=l$ we considered a related problem with initial data on the $t=0$ surface of the form \eqref{spatial2} near spatial infinity. It corresponds to a term proportional to the static moment. Equations \eqref{main1} and \eqref{main2} are the contributions to the late time tails due to these terms.  From the Couch-Torrence symmetry, it follows that such a term, if present near the bifurcation surface, will contribute to the $v^{-l-1}$ tail near the horizon. It seems likely that the iterative scheme of \cite{Barack} can be adopted in the frequency domain to compute tail contributions from $k < l$ terms.  

In section \ref{prelims} using the Couch-Torrence symmetry we also related higher multipole Aretakis and Newman-Penrose constants for a massless scalar in an extreme Reissner-Nordstr\"om black hole background.  Although a number of relations involving Pascal matrices are known in the literature, the identities \eqref{diagonalt} and \eqref{diagonaln} seem to be new.  We used these matrix relations to explain relations of functions, but from a mathematical perspective it would be more interesting to turn the logic around.  Namely, one could seek interpretations, e.g., through combinatorics or functional methods, of these and more general matrix relations.

All of our  analysis is only valid in the asymptotic regions, either near infinity or near the horizon $|r_*| \gg M$. 
We have not attempted to compute the correct radial dependence of the coefficient of the tail in full generality. From general results in the literature, we do expect the correct radial dependence of the tail at timelike infinfity to be the static solution to the extreme Reissner-Nordstr\"om potential \cite{Price:1971fb, Ori, Sela}, cf.~\eqref{decaying}
\be
\frac{r}{M} \left(\frac{r}{M}-1\right)^{-l-1}
\ee
with a constant pre-factor.  We expect that the constant pre-factor gets contributions from the Newman-Penrose constant as well as from the  Aretakis constant. This has been observed in numerical simulations \cite{Lucietti:2012xr}. The proportionality to the Aretakis constant is briefly discussed in \cite{Ori, Sela}, but details have not been presented. Together with the suggestion of references \cite{Winicour, Lucietti:2012xr} that ``initial static moments'' are more precisely thought of as initial data with non-zero Newman-Penrose constants, it is natural to conjecture that the total tail coefficient is proportional to the sum of (appropriately normalised) Aretakis and Newman-Penrose constants. It will be interesting to understand this circle of ideas better in the future. 

In a series of papers Casals, Gralla, and Zimmerman
\cite{Casals:2016mel, Zimmerman:2016qtn, Gralla:2018xzo} have analyzed the Aretakis instability and related questions in the frequency domain. They have obtained late time decay results on and off the horizon from the AdS$_2$ perspective. Their analysis is restricted to perturbations with vanishing Aretakis constants. When adapted to an extreme Reissner-Nordstr\"om black hole, and extended to perturbations with non-vanishing Aretakis constants, their analysis could be compared to ours through the Couch-Torrence duality.  It will be useful to relate our work to their work in detail. It will be very interesting to reproduce the late time tails from a microscopic CFT analysis for extreme black holes.

In a recent paper \cite{Camps}, Camps, Hadar, and Manton studied moduli space scattering of two extreme Reissner-Nordstr\"om black holes. They obtained the asymptotic gravitational radiation field wave-form at ``moderately'' late times, when the two black holes have not merged. They found that the asymptotic radiation field exhibits a quadrupolar late time tail of the form $t^{-2l-2}$ for $l=2$. It will be interesting to understand how their results relate to our analysis. We hope to report on some of these problems in our future work.

\subsection*{Acknowledgements} We thank Sayan Kar for discussions, and Shahar Hod and Orr Sela  for email correspondence.   Preliminary versions of these results were reported by AV at a ``Quantum Spacetime Seminar'' at TIFR, ``Indian Strings Meeting 2016'' at IISER Pune, and  ``IAGRG-2017'' conference at  IIT Guwahati. We thank Shiraz Minwalla and Julian Sonner for useful feedback. 
AV thanks BHU Varanasi for warm hospitality towards the final stages of this work.
PP thanks CMI Chennai for warm hospitality towards the final stages of this work. 
The work of AV is supported in part by the DST-Max Planck Partner Group ``Quantum Black Holes'' between CMI, Chennai and AEI Potsdam.
The work of SB is supported in part by the IIIT Allahabad seed grant ``Probing the interior of AdS Black Holes" and by DST Early Career Research Award ECR/2017/002124. BC and PP acknowledge hospitality at Kavli Asian Winter School ICTS/Prog-KAWS2018/01 and ICTP Trieste while this work was in progress.  The work of DDKC is supported by the Max Planck Society through the ``Gravitation and Black Hole Theory'' independent research group.  We also thank James Lucietti for reading through a version of the manuscript.

\end{document}